\documentclass[conference]{IEEEtran}
\IEEEoverridecommandlockouts
\usepackage{graphicx}
\usepackage{booktabs}
\usepackage{amsmath}
\usepackage{amssymb}
\usepackage{xcolor}
\usepackage{algorithm}
\usepackage{algorithmic}
\usepackage{subcaption}

\def\BibTeX{{\rm B\kern-.05em{\sc i\kern-.025em b}\kern-.08em
    T\kern-.1667em\lower.7ex\hbox{E}\kern-.125emX}}
    
\begin{document}

\title{Declarative Synthesis and Multi-Objective Optimization of Stripboard Circuit Layouts Using Answer Set Programming}

\author{\IEEEauthorblockN{Fang Li}
\IEEEauthorblockA{\textit{Computer Science Department} \\
\textit{Oklahoma Christian University}\\
Edmond, USA \\
fang.li@oc.edu}
}

\maketitle

\begin{abstract}
This paper presents a novel approach to automated stripboard circuit layout design using Answer Set Programming (ASP). The work formulates the layout problem as both a synthesis and multi-objective optimization task that simultaneously generates viable layouts while minimizing board area and component strip crossing. By leveraging ASP's declarative nature, this work expresses complex geometric and electrical constraints in a natural and concise manner. The two-phase solving methodology first ensures feasibility before optimizing layout quality. Experimental results demonstrate that this approach generates compact, manufacturable layouts for a range of circuit complexities. This work represents a significant advancement in automated stripboard layout, offering a practical tool for electronics prototyping and education while showcasing the power of declarative programming for solving complex design automation problems.
\end{abstract}

\begin{IEEEkeywords}
Electronic Design Automation, Circuit Layout, Answer Set Programming, Stripboard, Multi-Objective Optimization
\end{IEEEkeywords}

\section{Introduction}
Stripboard (also known as Veroboard) is widely used in electronics prototyping, offering a low-cost alternative to custom PCB fabrication. Despite its popularity, stripboard circuit layout remains a manual, time-consuming process where designers sketch layouts through trial and error. This contrasts sharply with the sophisticated automated tools available for PCB design.

This paper presents a novel approach using Answer Set Programming (ASP) to automate stripboard layout design. ASP's declarative nature allows for expressing complex geometric and electrical constraints naturally while leveraging optimized solving techniques.

Key contributions include:
\begin{enumerate}
    \item A formal ASP encoding of the stripboard layout problem
    \item A two-phase solving methodology separating feasibility from optimization
    \item A multi-objective optimization approach balancing board area and strip crossing minimization
    \item Experimental results demonstrating effectiveness across various circuit complexities
\end{enumerate}

\section{Background and Related Work}
\subsection{Stripboard Circuit Design and Automation}
Stripboard consists of a grid of holes with copper strips running in one direction. Components are inserted from the non-copper side, with connections made via continuous strips, while unwanted connections are prevented by cutting strips. Despite stripboard's popularity, there appears to be a significant gap in automation tools for generating optimized layouts from circuit schematics.

\subsection{Automated Circuit Layout}
PCB and IC layout automation typically involves two stages: component placement and connection routing. These NP-hard problems have been addressed using heuristic approaches like simulated annealing and genetic algorithms. However, stripboard layout differs fundamentally due to fixed routing channels and discrete hole placement.

\subsection{Answer Set Programming}

Answer Set Programming (ASP) is a declarative programming paradigm based on the stable model semantics of logic programming~\cite{Gelfond1988}. Problems are expressed as logical statements, with solutions corresponding to ``answer sets" satisfying all constraints. Modern ASP solvers like Clingo~\cite{Gebser2014} combine powerful techniques to solve complex problems effectively.

\subsection{Declarative Programming in Circuit Design}

Declarative programming approaches have been applied to various aspects of circuit design, from logical circuit analysis \cite{clocksin1987logic} to FPGA synthesis \cite{gill2011declarative} and image processing hardware \cite{ozkandeclarative}. This work extends these applications to physical stripboard layout synthesis, leveraging declarative programming's natural ability to handle complex constraints.

\section{Methods}
\label{sec:methods}

\begin{figure}[!t]
\centering
\includegraphics[width=0.4\textwidth]{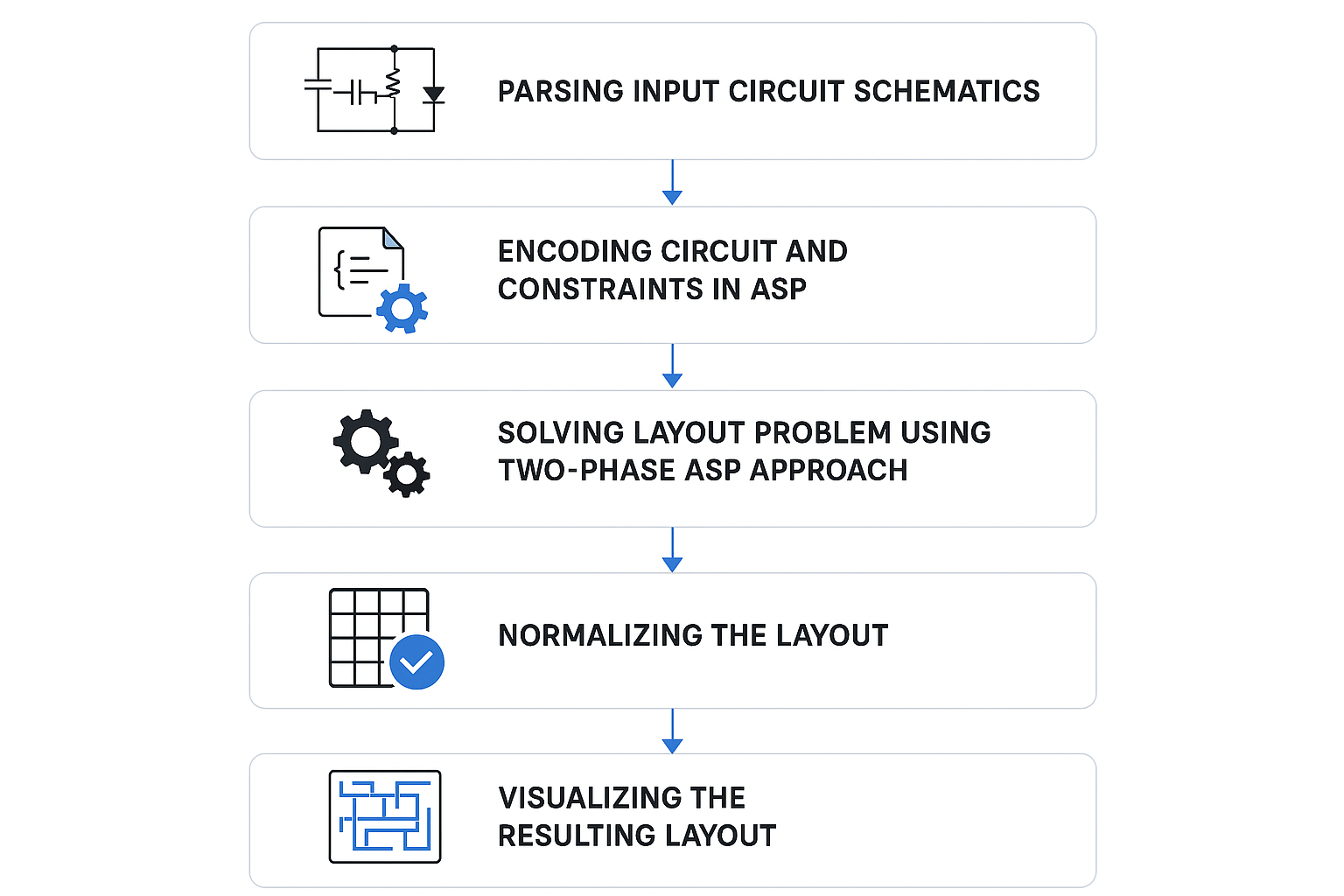}
\caption{Full workflow of the stripboard layout synthesis and optimization system}
\label{fig:workflow}
\end{figure}

This approach to stripboard layout synthesis and optimization involves a comprehensive workflow from circuit specification to final layout visualization. Figure~\ref{fig:workflow} illustrates this workflow.

\subsection{Input Processing}
The input to this system is a circuit netlist in KiCad v9 format. I developed a custom parser that extracts component information and connectivity from this format using a state machine approach. For each component, the parser identifies the reference, extracts component values, and determines component type. The connectivity analysis processes the netlist to identify each net and all component pins connected to that net. The output is a structured JSON representation containing detailed component information and electrical networks.

\subsection{ASP Encoding}
The JSON circuit representation is translated into ASP facts using a custom script. This translation generates ASP predicates that capture the circuit structure and serve as input to the ASP solver, including component definitions, pin configurations, and connectivity information.

The core ASP encoding models the stripboard layout problem using a 2D grid representation:

{\small\begin{verbatim} 
% Define dimensions of stripboard 
#const max_strips = 30. 
#const max_positions = 50.

strip(1..max_strips). 
position(1..max_positions).

% Each pin must be assigned exactly one position 
1 { pin_placement(Component, Pin, Strip, Pos)
: strip(Strip), position(Pos) } 1 
    :- component(Component, _, _), 
    pin(Component, Pin). 
\end{verbatim}}

This establishes the search space for the ASP solver, where each component pin must be assigned to exactly one location on the stripboard grid defined by a strip (row) and position (column).

\subsection{Two-Phase ASP Solving}
A key innovation in this approach is the use of a two-phase solving methodology that addresses the complexity of the synthesis and optimization problem.

\subsubsection{Phase 1: Feasibility Synthesis}
In the first phase, I invoke the Clingo ASP solver to find a feasible layout that satisfies all connectivity and geometric constraints without necessarily optimizing layout quality. Key constraints include:

{\small\begin{verbatim}
% Ensure electrical connectivity
:- circuit_net(C1, P1, NetworkId),
   circuit_net(C2, P2, NetworkId),
   pin_placement(C1, P1, S1, _),
   pin_placement(C2, P2, S2, _),
   S1 != S2.

% Prevent overlapping components 
:- pin_placement(C1, _, S, P),
   pin_placement(C2, _, S, P),
   C1 != C2.

% Ensure minimum distance for resistors
:- component(C, resistor, _),
   pin(C, 1), pin(C, 2),
   pin_placement(C, 1, _, P1),
   pin_placement(C, 2, _, P2),
   P2 - P1 < 3.

% Avoid component shorting
:- component(C, _, _),
   pin(C, 1),
   pin(C, 2),
   pin_placement(C, 1, S, _),
   pin_placement(C, 2, S, _).
\end{verbatim}}

The output of this phase is an initial placement of components that serves as a starting point for optimization.

\subsubsection{Phase 2: Optimization}
The second phase takes the output from Phase 1 and applies additional optimization criteria to refine the layout:

{\small\begin{verbatim}
% Define the size metrics
max_strip(MaxS) :- MaxS = 
  #max { S : used_strip(S) }.
max_position(MaxP) :- MaxP = 
  #max { P : used_position(P) }.
min_strip(MinS) :- MinS = 
  #min { S : used_strip(S) }.
min_position(MinP) :- MinP = 
  #min { P : used_position(P) }.

% Calculate the size of the stripboard
board_width(W) :- max_strip(MaxS), 
  min_strip(MinS), W = MaxS - MinS + 1.
board_length(L) :- max_position(MaxP),
  min_position(MinP), L = MaxP - MinP + 1.
board_area(A) :- board_width(W), 
  board_length(L), A = W * L.

% Calculate strip crossing distance
strip_distance(C, Distance) :-
    component(C, _, _),
    pin_placement(C, 1, S1, _),
    pin_placement(C, 2, S2, _),
    Distance = |S1 - S2|.

total_strip_distance(TotalDistance) :-
    TotalDistance = #sum { Distance, 
      C : strip_distance(C, Distance) }.

% Optimization goals (in order of priority)
#minimize { TD@3 : total_strip_distance(TD) }. 
#minimize { A@2 : board_area(A) }.     
#minimize { W@1 : board_width(W) }. 
\end{verbatim}}

This phase refines the layout to minimize specific metrics. The optimization goals, in order of priority, are:
\begin{itemize}
    \item Minimize the total strip distance for all components (primary objective)
    \item Minimize the overall board area (secondary objective)
    \item Prefer narrower boards when possible (tertiary objective)
\end{itemize}

By separating feasibility synthesis from optimization, I significantly reduce the search space that the solver must explore in each phase, enabling more efficient solving of complex circuits.

\subsection{Layout Normalization and Visualization}
\begin{figure}[!t]
\centering
\includegraphics[width=0.30\textwidth]{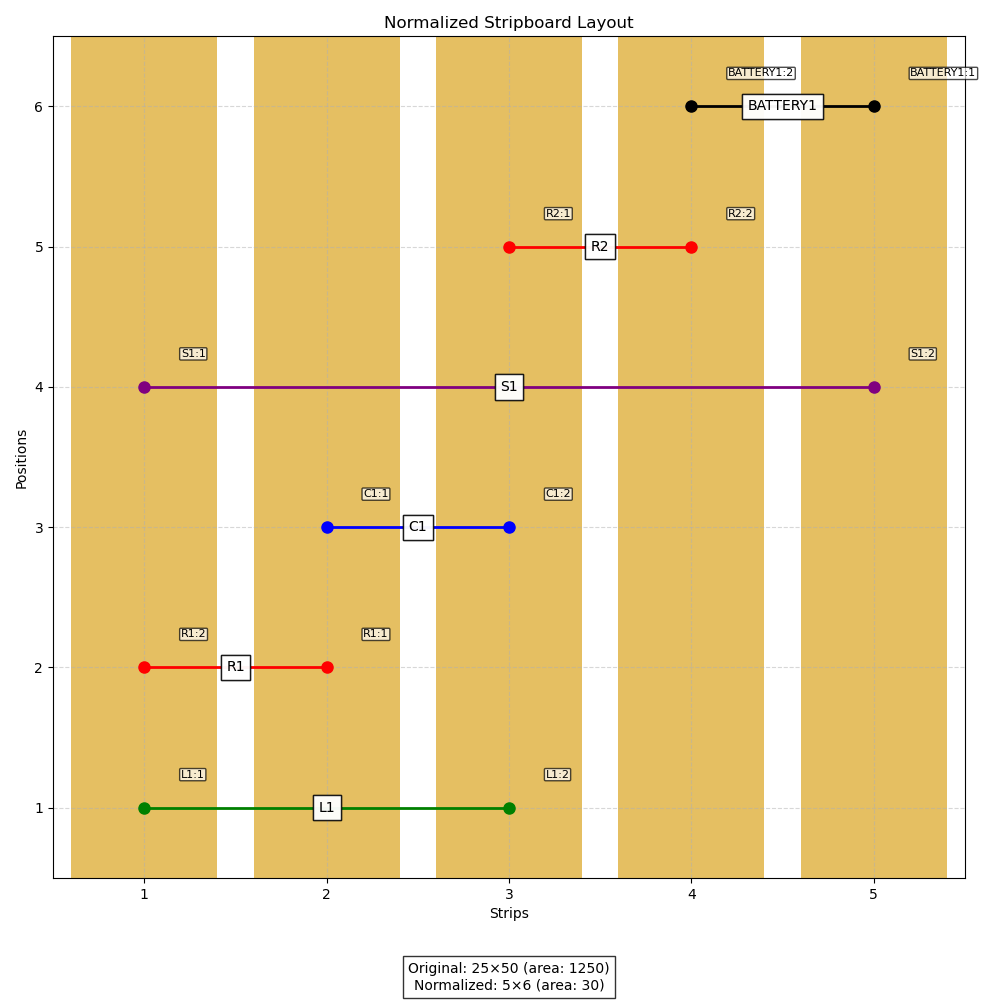}
\caption{Example visualization of a generated stripboard layout showing component placement, strips, and connections}
\label{fig:visualization}
\end{figure}

After optimization, I apply a normalization step to ensure the layout uses the minimum necessary board area. This script eliminates unnecessary gaps between components and shifts the entire layout to start at coordinates (0,0). Finally, I generate a visualization showing component placement, copper strips, cut locations, jumper wire routing, and board boundaries. Figure~\ref{fig:visualization} shows an example visualization of a generated layout.

\section{Results}
\subsection{Benchmark Circuits and Layout Quality}
To evaluate this approach, I tested the ASP-based layout synthesis and optimization system on a set of circuits of varying complexity:
\begin{itemize}
    \item LED Flasher (9 components, 7 networks)
    \item LRC (6 components, 5 networks)
    \item Op-Amp Filter (6 components, 5 networks)
    \item 4-bit Counter (18 components, 28 networks)
    \item Guitar Pedal (18 components, 12 networks)
\end{itemize}

Table~\ref{tab:results} presents the key metrics for the layouts generated by this system:

\begin{table}[!t]
\caption{Layout results for benchmark circuits}
\label{tab:results}
\centering
\small
\setlength{\tabcolsep}{3pt}
\begin{tabular}{|l|c|c|c|c|c|c|}
\hline
\textbf{Circuit} & \textbf{Board} & \textbf{Comp.} & \textbf{Net.} & \textbf{Phase 1} & \textbf{Phase 2} & \textbf{Total} \\
& \textbf{Size} & & & \textbf{(s)} & \textbf{(s)} & \textbf{(s)} \\ \hline
LED Flasher & 7×9 & 9 & 7 & 0.46 & 2.50 & 2.95 \\ \hline
LRC & 10×6 & 6 & 5 & 0.29 & 1.16 & 1.34 \\ \hline
Op-Amp Filter & 8×10 & 10 & 8 & 0.54 & 2.60 & 3.14 \\ \hline
4-bit Counter & 15×9 & 18 & 28 & 1.95 & 11.31 & 13.25 \\ \hline
Guitar Pedal & 12×18 & 18 & 12 & 1.52 & 10.40 & 11.92 \\ \hline
\end{tabular}
\end{table}

This approach successfully generated manufacturable layouts for all test circuits. Several key observations can be made from these results:
\begin{enumerate}
    \item \textbf{Compact Layouts}: The system consistently produced compact layouts with efficient use of board space. For example, the LED Flasher circuit was realized on a 7×9 grid (63 holes), which is remarkably compact for a 9-component circuit.
    
    \item \textbf{Component Placement}: The system effectively positioned components to optimize board area while maintaining sufficient space for manufacturability.
    
    \item \textbf{Computational Efficiency}: The two-phase approach allowed for efficient solving, with even the most complex circuit completing in about 13 seconds.
\end{enumerate}

\begin{figure}[!t]
\centering
\begin{subfigure}[b]{0.58\columnwidth}
    \centering
    \includegraphics[width=\textwidth]{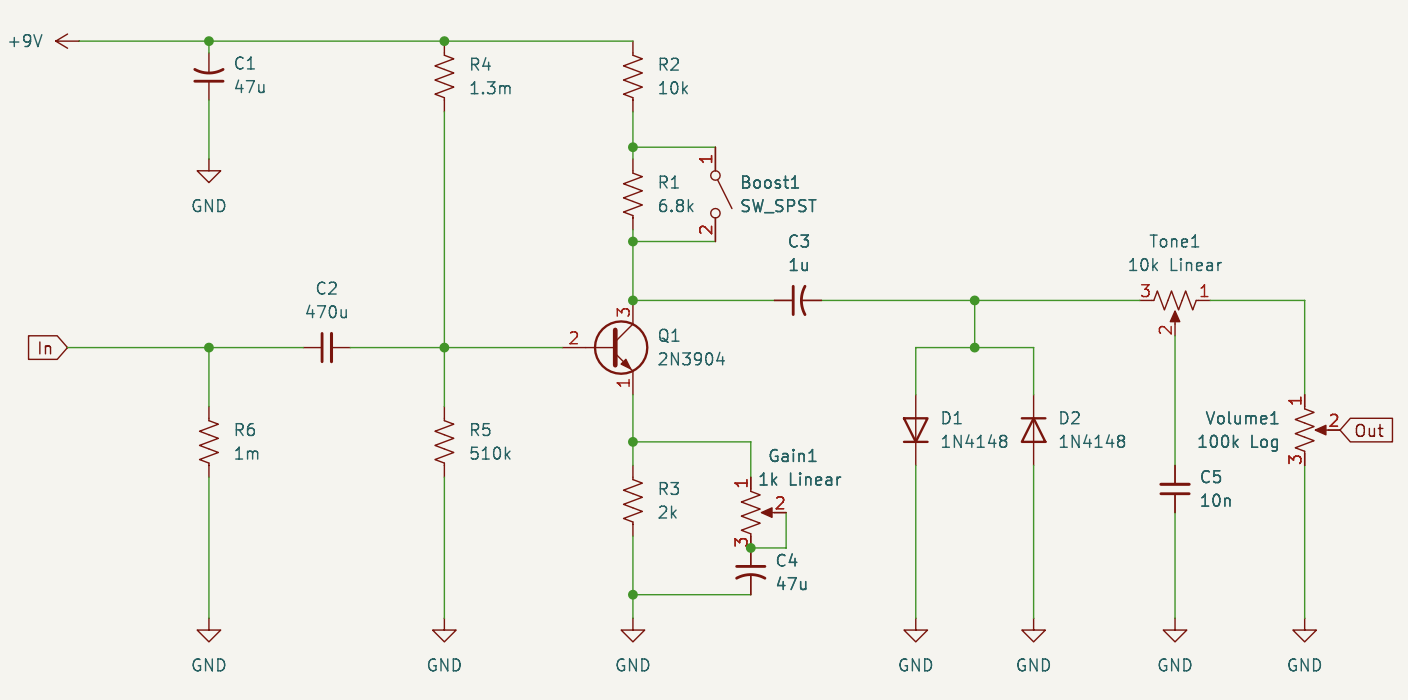}
    \caption{Original circuit schematic}
    \label{fig:pedal-schematic}
\end{subfigure}
\begin{subfigure}[b]{0.58\columnwidth}
    \centering
    \includegraphics[width=\textwidth]{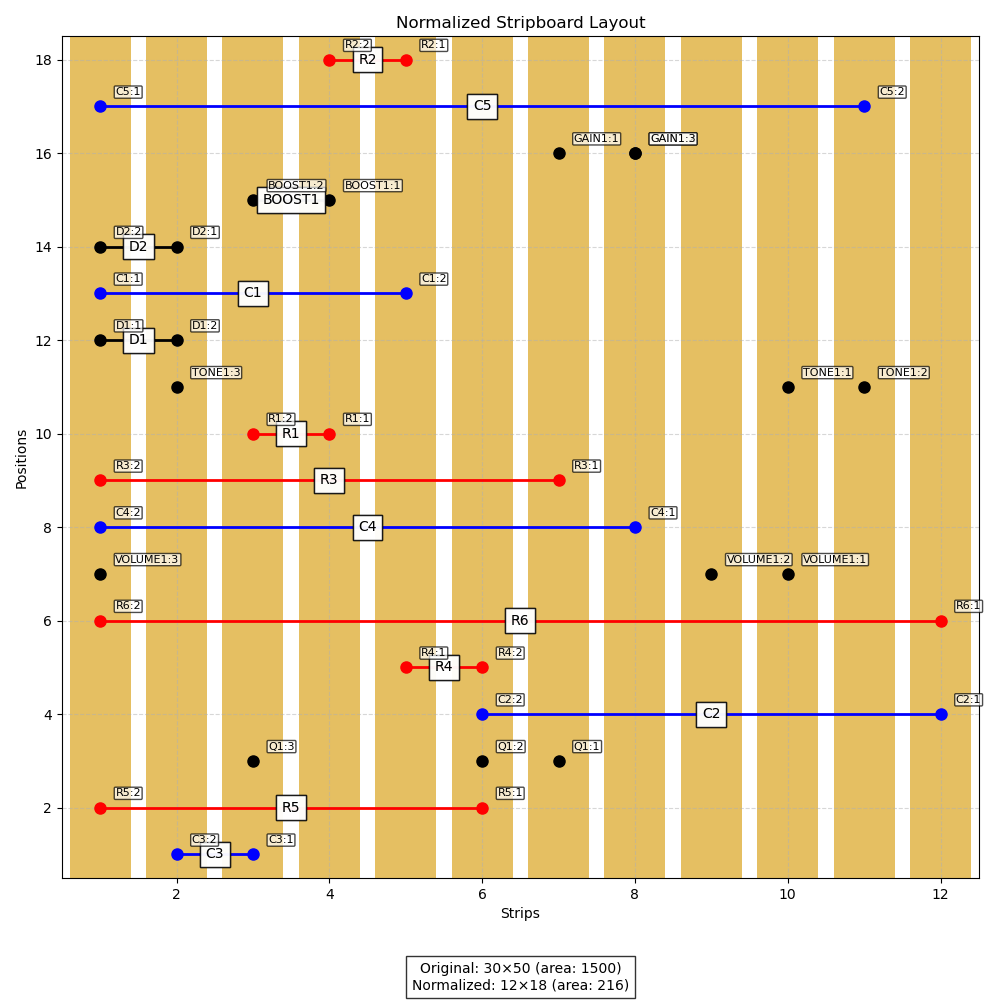}
    \caption{Generated stripboard layout}
    \label{fig:pedal-layout}
\end{subfigure}
\caption{Comparison between the original Guitar Pedal circuit schematic and the generated stripboard layout}
\label{fig:pedal}
\end{figure}

Figure~\ref{fig:pedal} shows a comparison between the original circuit schematic and the visualization of the generated Guitar Pedal layout. The layout demonstrates how the optimizer effectively places components to achieve a compact board design based on the original circuit design.

\subsection{Layout Verification}
To ensure the correctness of the generated layouts, I implemented a verification system that validates the electrical and physical properties of the generated solutions. The verification process consists of four key checks:
\begin{enumerate}
    \item \textbf{Electrical Connectivity Preservation}: The system extracts and compares electrical networks from both the original circuit netlist and the generated layout to ensure all connections are preserved.
    
    \item \textbf{Component Pin Placement}: The verification confirms that components with multiple pins are not placed with all pins on the same strip, which is crucial for components like resistors and capacitors.
    
    \item \textbf{Board Dimension Verification}: The system confirms that the reported board dimensions match the actual space used by components.
    
    \item \textbf{Layout Feasibility}: The verification checks that no pins overlap and that all pins are placed within the reported board dimensions.
\end{enumerate}

All generated layouts passed all verification checks, confirming that this ASP-based synthesis and optimization approach correctly preserves the electrical connections of the original circuits while producing compact, physically valid layouts.

\subsection{Effectiveness of Two-Phase Approach}
To evaluate the effectiveness of this two-phase approach, I compared its performance against a single-phase approach that combines the constraints and optimization goals into a single ASP program. Table~\ref{tab:two_phase_comparison} presents the results:

\begin{table}[!t]
\caption{Solving time comparison: Two-Phase vs. One-Phase Approach}
\label{tab:two_phase_comparison}
\centering
\setlength{\tabcolsep}{3pt}
\small
\begin{tabular}{|l|c|c|}
\hline
\textbf{Circuit} & \textbf{Two-Phase} & \textbf{One-Phase} \\
& \textbf{Total Time (s)} & \textbf{Total Time (s)} \\ \hline
LED Flasher     & 2.954 & 3.595 \\ \hline
LRC             & 1.486 & 1.501 \\ \hline
Op-Amp Filter   & 3.136 & 3.414 \\ \hline
4-bit Counter   & 13.254 & 17.812 \\ \hline
Guitar Pedal    & 11.924 & 15.491 \\ \hline
\end{tabular}
\end{table}

Based on these experimental results, the two-phase approach does not consistently provide a significant speedup for simpler circuits, but shows better performance for more complex designs like the 4-bit Counter and Guitar Pedal. The theoretical benefits of the two-phase approach are expected to be more pronounced for significantly larger or more complex circuits, where a single-phase approach would face a much larger search space and potentially become computationally intractable.

\subsection{Practical Applicability}
To assess the practical applicability of this approach, I physically constructed the Guitar Pedal circuit following the generated layout. This demonstrates the practical utility of this approach. The physical implementation confirmed that:
\begin{enumerate}
    \item The layout was manufacturable with standard stripboard techniques
    \item All electrical connections functioned as intended
    \item The layout was compact and efficient in its use of board space
\end{enumerate}

An additional benefit observed during physical construction was the clarity of the generated layout documentation, which eliminated ambiguity and reduced the likelihood of construction errors.

\section{Discussion}
\subsection{Strengths of the Declarative Approach}
This ASP-based approach offers several advantages:
\begin{enumerate}
    \item Natural expression of constraints with easy modification
    \item Complete exploration of the design space
    \item Clean separation of problem specification from solving strategy
    \item Flexible multi-objective optimization
\end{enumerate}

\subsection{Limitations and Future Work}
Despite its strengths, this approach has several limitations:
\begin{enumerate}
    \item Computational complexity for very large circuits
    \item Current focus on component placement without explicit strip cut optimization
    \item Limited component rotation options
    \item Opportunity for interactive synthesis combining user input with automation
\end{enumerate}

Future work will extend this approach to include strip cut placement and jumper wire routing optimization, and potentially address the broader domain of PCB layout.

\subsection{Conclusion}
This paper presented a novel ASP-based approach to stripboard circuit layout design. By formulating the problem as both synthesis and multi-objective optimization with a two-phase methodology, I demonstrated that declarative programming can produce high-quality, manufacturable layouts for electronic circuits.

This approach addresses the absence of automation tools for stripboard layout, providing what appears to be the first comprehensive solution for automated stripboard layout synthesis and optimization. Results confirm that the system generates compact, functional layouts while effectively managing computational complexity.

This work contributes not only a practical tool for stripboard design but also demonstrates the potential of declarative programming for electronic design automation as circuit complexity continues to increase.

\bibliographystyle{splncs04}
\bibliography{references}

\end{document}